\documentclass[aps,prb,twocolumn]{revtex4}
\usepackage{bm}
\usepackage{tabularx}
\usepackage{ascmac}
\usepackage{amsmath}
\usepackage{float}
\usepackage{wrapfig}
\usepackage{graphicx}
\usepackage[FIGBOTCAP]{subfigure}
\usepackage{color}

\bmdefine{\boldb}{b}
\bmdefine{\bolds}{s}
\bmdefine{\boldS}{S}
\bmdefine{\boldi}{i}
\bmdefine{\boldj}{j}
\bmdefine{\boldl}{l}
\bmdefine{\boldH}{H}
\bmdefine{\boldJ}{J}
\bmdefine{\boldx}{x}
\bmdefine{\boldX}{X}
\bmdefine{\boldk}{k}
\bmdefine{\boldK}{K}
\bmdefine{\boldp}{p}
\bmdefine{\bolde}{e}
\bmdefine{\boldq}{q}
\bmdefine{\boldQ}{Q}
\bmdefine{\boldD}{D}
\bmdefine{\boldr}{r}
\bmdefine{\boldR}{R}
\bmdefine{\boldn}{n}
\bmdefine{\boldv}{v}
\bmdefine{\boldA}{A}
\bmdefine{\boldzero}{0}

\begin{document}


\title{
Negative magneto-thermal-resistance 
in a disordered two-dimensional antiferromagnet
}


\author{Naoya Arakawa}
\email{naoya.arakawa@sci.toho-u.ac.jp} 
\author{Jun-ichiro Ohe}
\affiliation{
Department of Physics, Toho University, 
Funabashi, Chiba, 274-8510, Japan}


\begin{abstract}
We demonstrate that 
a weak external magnetic field can induce 
negative magneto-thermal-resistance for magnons 
in a disordered two-dimensional antiferromagnet. 
We study the main effect of a weak external magnetic field 
on the longitudinal thermal conductivity, $\kappa_{xx}$, 
for a disordered antiferromagnet 
using the weak-localization theory for magnons. 
We show that 
the weak-localization correction term of $\kappa_{xx}$ 
positively increases with increasing the magnetic field parallel to the ordered spins. 
Since this increase corresponds to a decrease of the thermal resistivity, 
this phenomenon is negative magneto-thermal-resistance for magnons. 
This negative magneto-thermal-resistance and the weak localization of magnons 
will be used to control the magnon thermal current 
in antiferromagnetic spintronics devices. 
We also discuss several implications for further experimental and theoretical
studies for disordered magnets.

\end{abstract}

\date{\today}
\maketitle


\section{Introduction}

Negative magnetoresistance can occur in a disordered electron system 
with a weak magnetic field. 
For electron systems without disorder, 
the resistivity increases as the magnetic field increases~\cite{Ziman}. 
This tendency is called positive magnetoresistance. 
If an electron system has impurities, 
the resistivity can decrease with increasing the magnetic field~\cite{Bergmann,
Nagaoka,Hikami,Maekawa}. 
This negative magnetoresistance is observed 
in a disordered two-dimensional electron system~\cite{exp-NegativeMR}. 

The above negative magnetoresistance 
originates from 
an effect of the magnetic field on the weak localization. 
In two dimensions, 
impurities can induce the weak localization of electrons~\cite{RG-4persons}, 
resulting in, for example, 
drastic suppression of the electron charge current 
parallel to an external electric field. 
This arises from 
the critical back scattering of electrons 
due to the multiple impurity scattering between electrons 
in the presence of time-reversal symmetry~\cite{Bergmann,Nagaoka}. 
Since the magnetic field breaks time-reversal symmetry, 
the magnetic field interferes with the weak localization~\cite{Hikami,Maekawa}. 
This effect results in a reduction in the resistivity. 

\begin{figure}[tb]
\includegraphics[width=36mm]{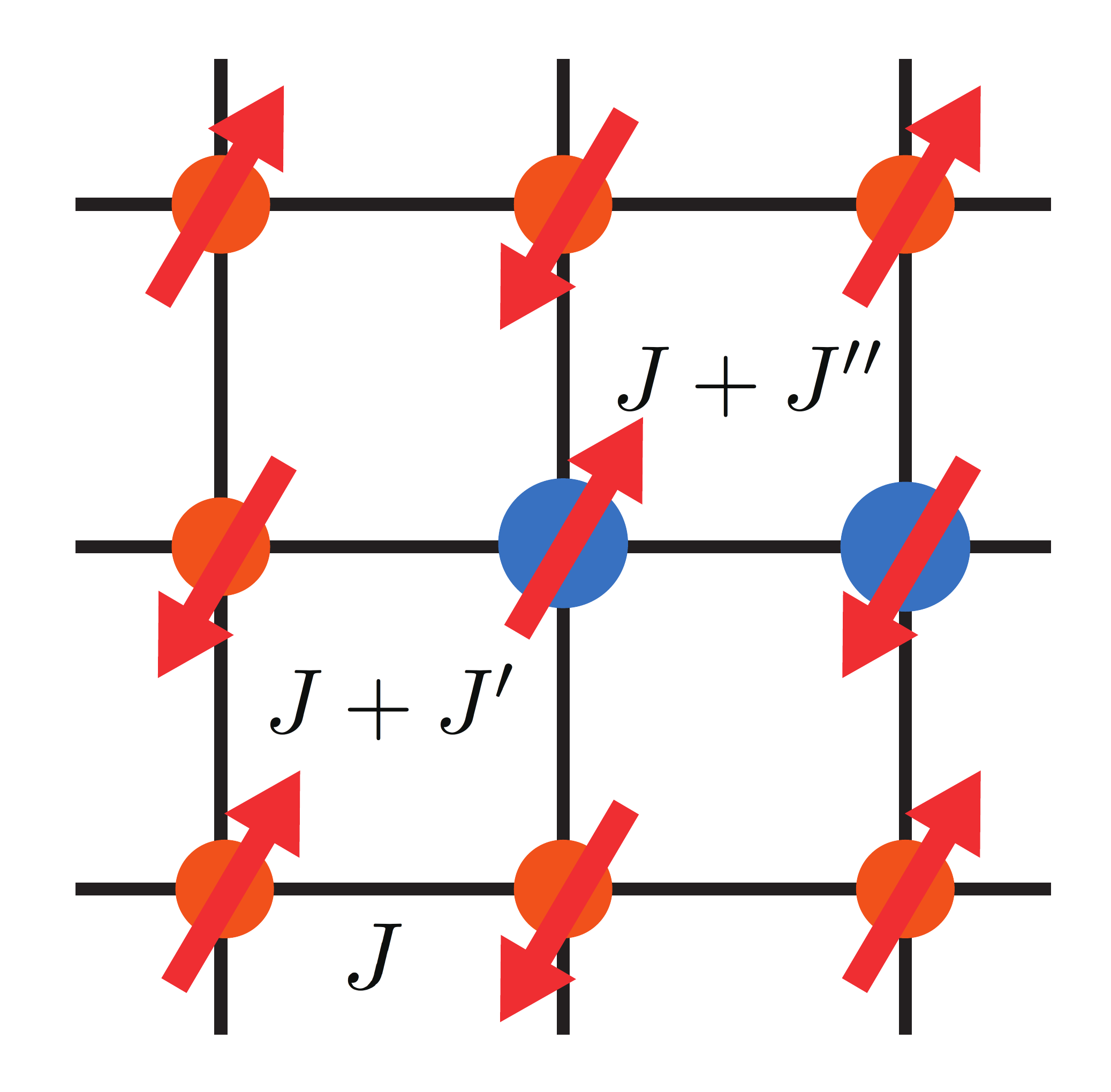}
\caption{
Schematic picture of our disordered two-dimensional antiferromagnet. 
Orange circles represent magnetic ions that exist even in the nondisordered system, 
and blue circles represent different magnetic ions. 
Up and down arrows represent spin-up and spin-down, respectively. 
The Heisenberg interactions between orangle circles, between an orange and a blue circle, 
and between blue circles 
are $J$, $J+J^{\prime}$, and $J+J^{\prime\prime}$, respectively. 
For more details, see Sec. II and Appendix A. 
}
\label{fig1}
\end{figure}

A similar magneto-transport phenomenon may occur 
in a disordered antiferromagnet 
with a weak external magnetic field. 
In a disordered two-dimensional antiferromagnet (Fig. \ref{fig1}), 
the critical back scattering of magnons 
drastically suppresses the magnon thermal current parallel to 
temperature gradient~\cite{Lett}. 
This is the weak localization of magnons. 
Since antiferromagnets have time-reversal symmetry, 
the effect of an external magnetic field may lead to 
a magneto-thermal-transport phenomenon characteristic of the disordered magnets. 

In this paper, 
we study the longitudinal thermal conductivity, $\kappa_{xx}$, 
for a disordered antiferromagnet with a weak external magnetic field. 
As an effective model, 
we use the Hamiltonian, which consists of 
the antiferromagnetic Heisenberg interaction and magnetic anisotropy, 
the mean-field type impurity potential, 
and the Zeeman coupling. 
Extending the weak-localization theory~\cite{Lett} 
for a disordered antiferromagnet to the case with the weak magnetic field, 
we analyze its main effect on $\kappa_{xx}$. 
We show that 
as the magnetic field increases, 
$\kappa_{xx}$ increases due to the positive increase of 
the weak-localization correction term of $\kappa_{xx}$ 
in a similar way to the negative magnetoresistance 
for electrons~\cite{Nagaoka,Bergmann,Hikami}. 
This is negative magneto-thermal-resistance for magnons 
due to the effects of the weak localization and the weak magnetic field. 
Then, we discuss the similarities and differences 
between our phenomenon and the electrons' phenomenon, 
and provide experimental and theoretical implications. 
Throughout this paper, we set $\hbar=1$ and $k_{\textrm{B}}=1$. 

\section{Model}

Our Hamiltonian consists of three parts as follows:
\begin{align}
\hat{H}=\hat{H}_{0}+\hat{H}_{\textrm{imp}}+\hat{H}_{\textrm{Z}}.\label{eq:H}
\end{align}
Here $\hat{H}_{0}$ is the Hamiltonian of an antiferromagnet without impurities, 
$\hat{H}_{\textrm{imp}}$ is the impurity Hamiltonian, 
and $\hat{H}_{\textrm{Z}}$ is the Hamiltonian of an external magnetic field. 
First, 
$\hat{H}_{0}$ is given by 
the nearest-neighbor antiferromagnetic Heisenberg interaction 
and the magnetic anisotropy as follows:
\begin{align}
\hat{H}_{0}
=2J\sum\limits_{\langle \boldi,\boldj\rangle}\hat{\boldS}_{\boldi}\cdot \hat{\boldS}_{\boldj}
-K\Bigl[\sum\limits_{\boldi\in A}(\hat{S}^{z}_{\boldi})^{2}
+\sum\limits_{\boldj\in B}(\hat{S}^{z}_{\boldj})^{2}\Bigr],\label{eq:H0-S}
\end{align}
where site indices $\boldi$ and $\boldj$ satisfy 
$\boldi\in A$ and $\boldj\in B$ 
for $A$ or $B$ sublattice. 
We have considered the positive $J$ and $K$. 
Second, 
$\hat{H}_{\textrm{imp}}$ is given by the mean-field-type impurity potential~\cite{Lett} as follows:
\begin{align}
\hat{H}_{\textrm{imp}}
=-\sum\limits_{\boldi\in A_{\textrm{imp}}}
V_{\textrm{imp}}\hat{S}_{\boldi}^{z}
+\sum\limits_{\boldj\in B_{\textrm{imp}}}
V_{\textrm{imp}}\hat{S}_{\boldj}^{z}.\label{eq:Himp-S}
\end{align}  
This Hamiltonian describes the main effect of impurities, i.e., 
the change of the exchange interaction 
due to substituting part of magnetic ions by different magnetic ions~\cite{Lett}; 
we treat this partial substitution as 
randomly distributed impurities for magnets~\cite{Lett} (see Fig. \ref{fig1}). 
For more details, see Appendix A. 
We suppose that 
the numbers of $A_{\textrm{imp}}$ and $B_{\textrm{imp}}$ are the same. 
Third, 
$\hat{H}_{\textrm{Z}}$ is given by the Zeeman coupling as follows:
\begin{align}
\hat{H}_{\textrm{Z}}
=-H\sum\limits_{\boldi\in A}
\hat{S}_{\boldi}^{z}
-H\sum\limits_{\boldj\in B}
\hat{S}_{\boldj}^{z}.\label{eq:Hz-S}
\end{align}

Then, 
we can express our Hamiltonian 
in terms of magnon operators using the linear-spin-wave approximation~\cite{Anderson-SW} 
for a collinear antiferromagnet. 
Using it, we obtain
\begin{align}
\hat{H}_{0}=&
\sum\limits_{\boldq}
\sum\limits_{l,l^{\prime}=A,B}\epsilon_{ll^{\prime}}(\boldq)
\hat{x}^{\dagger}_{\boldq l}\hat{x}_{\boldq l^{\prime}},\label{eq:H0-LSWA}\\
\hat{H}_{\textrm{imp}}=&
\sum\limits_{\boldq,\boldq^{\prime}}
\sum\limits_{l=A,B}
V^{\textrm{imp}}_{l}(\boldq-\boldq^{\prime})
\hat{x}_{\boldq l}^{\dagger}\hat{x}_{\boldq^{\prime} l},\label{eq:Himp-LSWA}\\
\hat{H}_{\textrm{Z}}=&
\sum\limits_{\boldq}
\sum\limits_{l=A,B}
H_{l}
\hat{x}_{\boldq l}^{\dagger}\hat{x}_{\boldq l}.\label{eq:Hz-LSWA}
\end{align}
Each quantity in those equations is defined as follows. 
$\epsilon_{ll^{\prime}}(\boldq)$ is given by
\begin{align}
\epsilon_{ll^{\prime}}(\boldq)=
\begin{cases}
2S[J(\boldzero)+K]\ \ \ (l=l^{\prime})\\
2SJ(\boldq)\ \ \ \ \ \ \ \ \ \ \ (l\neq l^{\prime})
\end{cases},
\end{align}
where $S$ is spin quantum number, 
and $J(\boldq)=J\sum_{j=1}^{z}e^{i\boldq\cdot \boldr_{j}}$ 
with $z$, coordination number. 
Magnon operators $\hat{x}_{\boldq l}$ and $\hat{x}_{\boldq l}^{\dagger}$ are given by 
\begin{align}
&\hat{x}_{\boldq l}=
\begin{cases}
\hat{a}_{\boldq} \ \ \ \ (l=A) \\
\hat{b}_{\boldq}^{\dagger} \ \ \ \ (l=B)
\end{cases},
\end{align}
and
\begin{align}
&\hat{x}_{\boldq l}^{\dagger}=
\begin{cases}
\hat{a}_{\boldq}^{\dagger} \ \ \ \ (l=A) \\
\hat{b}_{\boldq} \ \ \ \ (l=B)
\end{cases},
\end{align}
where $\hat{a}_{\boldq}$ and $\hat{a}_{\boldq}^{\dagger}$ are 
annihilation and creation operators of a magnon for $A$ sublattice, 
and $\hat{b}_{\boldq}$ and $\hat{b}_{\boldq}^{\dagger}$ are those for $B$ sublattice.  
$V^{\textrm{imp}}_{l}(\boldQ)$ is given by
\begin{align}
V^{\textrm{imp}}_{l}(\boldQ)=
\begin{cases}
V_{\textrm{imp}}\dfrac{2}{N}\sum\limits_{\boldi\in A_{\textrm{imp}}}e^{i\boldQ\cdot \boldi}\ \ \ \ \ (l=A)\\
V_{\textrm{imp}}\dfrac{2}{N}\sum\limits_{\boldj\in B_{\textrm{imp}}}e^{i\boldQ\cdot \boldj}\ \ \ \ \ (l=B)
\end{cases},\label{eq:Vimp-q}
\end{align} 
where $N$ is the total number of sites. 
Note that due to the restriction of the sum of sites in $\hat{H}_{\textrm{imp}}$ 
[see Eqs. (\ref{eq:Himp-S}) and (\ref{eq:Vimp-q})], 
$\hat{H}_{\textrm{imp}}$ is non-diagonal in terms of momentum, 
as seen from Eq. (\ref{eq:Himp-LSWA}). 
This property is the origin of the finite back scattering in disordered systems; 
however, the finite back scattering does not always imply the localization of quasiparticles, 
such as magnons. 
$H_{l}$ is given by
\begin{align}
H_{l}=
\begin{cases}
H\ \ \ \ \ (l=A)\\
-H \ \ \ (l=B)
\end{cases}.
\end{align}

We can also rewrite our Hamiltonian 
in the band representation 
using the Bogoliubov transformation~\cite{Anderson-SW}, 
\begin{align}
\hat{x}_{\boldq l}=\sum\limits_{\nu=\alpha,\beta}U_{l\nu}(\boldq)\hat{x}_{\boldq \nu}.
\end{align}
The transformation matrix $U_{l\nu}(\boldq)$ 
is so determined that 
the matrix of $\hat{H}_{0}+\hat{H}_{\textrm{Z}}$ is diagonalized. 
We thus get 
\begin{align}
&U_{A\alpha}(\boldq)=U_{B\beta}(\boldq)=\cosh \theta_{\boldq},\\
&U_{A\beta}(\boldq)=U_{B\alpha}(\boldq)=-\sinh \theta_{\boldq},
\end{align}
where the hyperbolic functions satisfy 
\begin{align}
\tanh 2\theta_{\boldq}=\dfrac{\epsilon_{AB}(\boldq)}{\epsilon_{AA}(\boldq)}.
\end{align} 
As a result of the diagonalization, 
we obtain 
\begin{align}
\hat{H}_{0}+\hat{H}_{\textrm{Z}}
=\sum\limits_{\boldq}\sum\limits_{\nu=\alpha,\beta}\epsilon_{\boldq \nu}
\hat{x}_{\boldq \nu}^{\dagger}\hat{x}_{\boldq \nu},
\end{align}
and 
\begin{align}
\epsilon_{\boldq \nu}=
\begin{cases}
\epsilon_{\boldq}+H \ \ \ (\nu=\alpha)\\
\epsilon_{\boldq}-H \ \ \ (\nu=\beta)
\end{cases},\label{eq:magnon-energy}
\end{align}
where $\epsilon_{\boldq}=\sqrt{\epsilon_{AA}(\boldq)^{2}-\epsilon_{AB}(\boldq)^{2}}$.
Since the magnon energy should be non-negative, 
the external magnetic field should be smaller than the magnon dispersion energy 
for discussions about the effect of the external magnetic field 
on magnon transport of antiferromagnets. 
This is the reason why we consider only the weak-field case of the external magnetic field. 

\section{Magneto-thermal-transport}

As a magneto-transport property, 
we consider 
the longitudinal thermal conductivity $\kappa_{xx}$ 
under the assumptions of local equilibrium and local energy conservation. 
$\kappa_{xx}$ is given by $j_{Q}^{x}=\kappa_{xx}(-\partial_{x} T)$, where  
$(-\partial_{x} T)$ is temperature gradient and  
$j_{Q}^{x}$ is the thermal current. 
Since the magnon thermal current is equal to 
the magnon energy current because of no charge current, 
we use the thermal current and the energy current for magnons in the same sense. 
Due to local energy conservation, 
we can derive the magnon energy current for our model 
in a similar way to the electron charge current~\cite{Mahan-text,Lett}: 
the energy current operator is determined by~\cite{Mahan-text} 
\begin{align}
\hat{\boldJ}_{\textrm{E}}=
i\sum\limits_{\boldi,\boldj}
\boldr_{\boldi}[\hat{h}_{\boldj},\hat{h}_{\boldi}],\label{eq:def-JE}
\end{align}
where $\hat{h}_{\boldi}$ is defined by 
$\hat{H}=\sum\textstyle_{\boldi}\hat{h}_{\boldi}$. 
By calculating the right-hand side of Eq. (\ref{eq:def-JE}) for our model, 
we obtain 
the energy current operator,
\begin{align}
\hat{\boldJ}_{\textrm{E}}
=&\sum\limits_{\boldq}\sum\limits_{l,l^{\prime}=A,B}
\bolde_{ll^{\prime}}(\boldq)\hat{x}^{\dagger}_{\boldq l}\hat{x}_{\boldq l^{\prime}}\notag\\
=&
\sum\limits_{\boldq}
\epsilon_{AB}(\boldq)
\frac{\partial \epsilon_{AB}(\boldq)}{\partial \boldq}
(\hat{a}^{\dagger}_{\boldq}\hat{a}_{\boldq}
-\hat{b}_{\boldq}\hat{b}^{\dagger}_{\boldq}).\label{eq:JE}
\end{align}
Here, $\bolde_{ll^{\prime}}(\boldq)$ has been defined as 
$\bolde_{AA}(\boldq)=-\bolde_{BB}(\boldq)=
\epsilon_{AB}(\boldq)
\frac{\partial \epsilon_{AB}(\boldq)}{\partial \boldq}$ 
and $\bolde_{AB}(\boldq)=\bolde_{BA}(\boldq)=\boldzero$. 

In the weak-localization regime, 
we can express $\kappa_{xx}$ as~\cite{Lett}
\begin{align}
\kappa_{xx}=\kappa_{xx}^{(\textrm{Born})}+\Delta\kappa_{xx},\label{eq:kappa-sum}
\end{align}
where $\kappa_{xx}^{(\textrm{Born})}$ is the longitudinal thermal conductivity 
in the Born approximation, 
\begin{align}
\kappa_{xx}^{(\textrm{Born})}=&\frac{1}{TN}
\sum\limits_{\boldq}
\sum\limits_{l_{1},l_{2},l_{3},l_{4}}
e_{l_{1}l_{2}}^{x}(\boldq)e_{l_{3}l_{4}}^{x}(\boldq)
P\hspace{-4pt}\int^{\infty}_{-\infty}\hspace{-2pt}\frac{d\epsilon}{2\pi}
\notag\\
&\times 
\Bigl[-\frac{\partial n(\epsilon)}{\partial \epsilon}\Bigr]
\bar{D}_{l_{4}l_{1}}^{(\textrm{A})}(\boldq,\epsilon)
\bar{D}_{l_{2}l_{3}}^{(\textrm{R})}(\boldq,\epsilon),\label{eq:kappa^Born}
\end{align}
and $\Delta\kappa_{xx}$ is the weak-localization correction term, 
\begin{align}
\Delta\kappa_{xx}
=&\frac{1}{TN}
\sum\limits_{\boldq,\boldq^{\prime}}
\sum\limits_{l_{1},l_{2},l_{3},l_{4}}
e_{l_{1}l_{2}}^{x}(\boldq)e_{l_{3}l_{4}}^{x}(\boldq^{\prime})
P\hspace{-4pt}\int^{\infty}_{-\infty}\hspace{-2pt}\frac{d\epsilon}{2\pi}
\notag\\
&\times 
\Bigl[-\frac{\partial n(\epsilon)}{\partial \epsilon}\Bigr]
\sum\limits_{l,l^{\prime}}
\bar{D}_{l_{4}l^{\prime}}^{(\textrm{A})}(\boldq^{\prime},\epsilon)
\bar{D}_{ll_{3}}^{(\textrm{R})}(\boldq^{\prime},\epsilon)\notag\\
&\times 
\Gamma_{l^{\prime}l}(\boldq+\boldq^{\prime},\epsilon)
\bar{D}_{ll_{1}}^{(\textrm{A})}(\boldq,\epsilon)
\bar{D}_{l_{2}l^{\prime}}^{(\textrm{R})}(\boldq,\epsilon).\label{eq:kappa^VC}
\end{align}
For the derivation, see Appendix B. 
In those equations, 
$n(\epsilon)$ is the Bose distribution function, 
$n(\epsilon)=(e^{\epsilon/T}-1)^{-1}$; 
$\bar{D}_{ll^{\prime}}^{(\textrm{R})}(\boldq,\epsilon)$ 
and $\bar{D}_{ll^{\prime}}^{(\textrm{A})}(\boldq,\epsilon)$ 
are retarded and advanced Green's functions of magnons after taking the impurity averaging; 
$\Gamma_{l^{\prime}l}(\boldq+\boldq^{\prime},\epsilon)$ 
is the particle-particle-type four-point vertex function 
due to the multiple impurity scattering. 
Furthermore, 
the vertex function and the Green's functions are connected 
by the Bethe-Salpeter equation,
\begin{align}
\Gamma_{ll^{\prime}}(\boldQ,\omega)
=&\gamma_{\textrm{imp}}\Pi_{ll^{\prime}}(\boldQ,\omega)\gamma_{\textrm{imp}}\notag\\
&+\sum\limits_{l^{\prime\prime}}
\gamma_{\textrm{imp}}\Pi_{ll^{\prime\prime}}(\boldQ,\omega)
\Gamma_{l^{\prime\prime}l^{\prime}}(\boldQ,\omega),\label{eq:BSeq}
\end{align}
where $\gamma_{\textrm{imp}}=\frac{2}{N}n_{\textrm{imp}}V_{\textrm{imp}}^{2}$ 
with the impurity concentration $n_{\textrm{imp}}$, and 
\begin{align}
\Pi_{ll^{\prime}}(\boldQ,\omega)
=\sum\limits_{\boldq_{1}}
\bar{D}_{ll^{\prime}}^{(\textrm{R})}(\boldq_{1},\omega)
\bar{D}_{ll^{\prime}}^{(\textrm{A})}(\boldQ-\boldq_{1},\omega).\label{eq:Pi}
\end{align}

To analyze the main effect of the weak magnetic field on $\kappa_{xx}$, 
we first analyze the magnon Green's functions. 
We can express 
the retarded Green's function in the absence of impurities as follows:
\begin{align}
\hspace{-10pt}
D_{ll^{\prime}}^{0(\textrm{R})}(\boldq,\omega)
=\dfrac{U_{l\alpha}(\boldq)U_{l^{\prime}\alpha}(\boldq)}{\omega-\epsilon_{\boldq}-H+i\delta}
-\dfrac{U_{l\beta}(\boldq)U_{l^{\prime}\beta}(\boldq)}{\omega+\epsilon_{\boldq}-H+i\delta},
\label{eq:D^R0}
\end{align}
where $\delta=0+$. 
Since for the weak magnetic field, 
$H$ is smaller than the magnon dispersion energy, 
the main contribution for $\omega >0$ comes from 
the first term of the right-hand side of Eq. (\ref{eq:D^R0}), 
the positive-pole contribution;  
the main contribution 
for $\omega < 0$ comes from the second term, the negative-pole contribution. 
We thus approximate $D_{ll^{\prime}}^{0(\textrm{R})}(\boldq,\omega)$ as
\begin{align}
D_{ll^{\prime}}^{0(\textrm{R})}(\boldq,\omega)\sim 
\begin{cases}
\dfrac{U_{l\alpha}(\boldq)U_{l^{\prime}\alpha}(\boldq)}{\omega-\epsilon_{\boldq}-H+i\delta}
\ \ \ \ (\omega > 0)\\[8pt]
-\dfrac{U_{l\beta}(\boldq)U_{l^{\prime}\beta}(\boldq)}{\omega+\epsilon_{\boldq}-H+i\delta}
\ \ (\omega < 0)
\end{cases}\hspace{-10pt}.\label{eq:D^R0-approx}
\end{align}
Replacing $\delta$ in Eq. (\ref{eq:D^R0-approx}) by $-\delta$,
we obtain $D_{ll^{\prime}}^{0(\textrm{A})}(\boldq,\omega)$. 
Then, 
we can derive the magnon Green's functions 
in the presence of impurities by using the Dyson equation 
and taking the impurity averaging~\cite{Lett}; 
the Dyson equation, for example, for retarded quantities is 
$\bar{D}_{ll^{\prime}}^{(\textrm{R})}(\boldq,\omega)
=D_{ll^{\prime}}^{0(\textrm{R})}(\boldq,\omega)
+\sum_{l^{\prime\prime}}
D_{ll^{\prime\prime}}^{0(\textrm{R})}(\boldq,\omega)
\Sigma_{l^{\prime\prime}}^{(\textrm{R})}(\omega)
\bar{D}_{l^{\prime\prime}l^{\prime}}^{(\textrm{R})}(\boldq,\omega)$
with the self-energy in the Born approximation.    
In that derivation, we neglect the real part of the self-energy 
and consider only its imaginary part 
because the imaginary part is vital 
for the weak localization~\cite{Bergmann,Nagaoka}. 
As a result, we obtain
\begin{align}
\hspace{-10pt}
&\bar{D}_{ll^{\prime}}^{(\textrm{R})}(\boldq,\omega)\notag\\
\hspace{-10pt}
\sim &
\begin{cases}
\dfrac{U_{l\alpha}(\boldq)U_{l^{\prime}\alpha}(\boldq)}
{\omega-\epsilon_{\boldq}-H+i[\tilde{\gamma}(\omega)+\tilde{\gamma}^{H}(\omega)]}
\ \ \ \ \ \ \ \  \ (\omega > 0)\\[8pt]
-\dfrac{U_{l\beta}(\boldq)U_{l^{\prime}\beta}(\boldq)}
{\omega+\epsilon_{\boldq}-H+i[\tilde{\gamma}(-\omega)+\tilde{\gamma}^{H}(-\omega)]}
\ \ (\omega < 0)
\end{cases}\hspace{-10pt}.\label{eq:D^R-pre}
\end{align}
Here $\tilde{\gamma}(\pm\omega)$ is the damping that is finite 
even for $H=0$,
\begin{align}
\tilde{\gamma}(\pm\omega)
=&(\cosh^{4}\theta_{\boldq}+\sinh^{4}\theta_{\boldq})\pi n_{\textrm{imp}}V_{\textrm{imp}}^{2}
\rho(\pm\omega)\notag\\
=&(\cosh^{4}\theta_{\boldq}+\sinh^{4}\theta_{\boldq})\gamma(\pm\omega),\label{eq:gam-nonH}
\end{align}
and $\tilde{\gamma}^{H}(\pm\omega)$ is the damping that is finite 
only for $H\neq 0$, 
\begin{align}
\tilde{\gamma}^{H}(\pm\omega)
=&(\cosh^{4}\theta_{\boldq}+\sinh^{4}\theta_{\boldq})\pi n_{\textrm{imp}}V_{\textrm{imp}}^{2}\notag\\
&\times 
[\rho(\pm\omega\mp H)-\rho(\pm\omega)]\notag\\
=&(\cosh^{4}\theta_{\boldq}+\sinh^{4}\theta_{\boldq})\gamma^{H}(\pm\omega),\label{eq:gam-H}
\end{align}
where $\rho(\omega)$ is the density of states for magnons, and  
$\boldq$ of $\cosh^{4}\theta_{\boldq}$ and $\sinh^{4}\theta_{\boldq}$ 
are determined by $\epsilon_{\boldq}=|\omega|$. 
For the sake of simplicity, 
we consider only the magnetic-field effect coming from the damping 
and neglect the other effect hereafter 
because 
the effect of the energy shifts in the denominators of Eq. (\ref{eq:D^R-pre}) 
is small for weak $H$ and is similar to the effect of the real part of the self-energy. 
As a result, 
$\bar{D}_{ll^{\prime}}^{(\textrm{R})}(\boldq,\omega)$ is expressed as follows:
\begin{align}
&\bar{D}_{ll^{\prime}}^{(\textrm{R})}(\boldq,\omega)\notag\\
=& 
\begin{cases}
\dfrac{U_{l\alpha}(\boldq)U_{l^{\prime}\alpha}(\boldq)}
{\omega-\epsilon_{\boldq}+i[\tilde{\gamma}(\omega)+\tilde{\gamma}^{H}(\omega)]}
\ \ \ \ \ \ \ \  \ (\omega > 0)\\[8pt]
-\dfrac{U_{l\beta}(\boldq)U_{l^{\prime}\beta}(\boldq)}
{\omega+\epsilon_{\boldq}+i[\tilde{\gamma}(-\omega)+\tilde{\gamma}^{H}(-\omega)]}
\ \ (\omega < 0)
\end{cases}\hspace{-10pt}.\label{eq:D^R}
\end{align}
Similarly, 
we can express $\bar{D}_{ll^{\prime}}^{(\textrm{A})}(\boldq,\omega)$ as follows:
\begin{align}
&\bar{D}_{ll^{\prime}}^{(\textrm{A})}(\boldq,\omega)\notag\\
=& 
\begin{cases}
\dfrac{U_{l\alpha}(\boldq)U_{l^{\prime}\alpha}(\boldq)}
{\omega-\epsilon_{\boldq}-i[\tilde{\gamma}(\omega)+\tilde{\gamma}^{H}(\omega)]}
\ \ \ \ \ \ \ \  \ (\omega > 0)\\[8pt]
-\dfrac{U_{l\beta}(\boldq)U_{l^{\prime}\beta}(\boldq)}
{\omega+\epsilon_{\boldq}-i[\tilde{\gamma}(-\omega)+\tilde{\gamma}^{H}(-\omega)]}
\ \ (\omega < 0)
\end{cases}\hspace{-10pt}.\label{eq:D^A}
\end{align}

We next analyze $\Pi_{ll^{\prime}}(\boldQ,\omega)$ and $\Gamma_{ll^{\prime}}(\boldQ,\omega)$ 
for small $Q=|\boldQ|$ in the weak magnetic field. 
By combining Eqs. (\ref{eq:D^R}) and (\ref{eq:D^A}) for $\omega > 0$ 
with Eq. (\ref{eq:Pi}), 
we have
\begin{align}
&\Pi_{ll^{\prime}}(\boldQ,\omega)
=\sum\limits_{\boldq_{1}}
\frac{U_{l\alpha}(\boldq_{1})U_{l^{\prime}\alpha}(\boldq_{1})}
{\omega-\epsilon_{\boldq_{1}}+i[\tilde{\gamma}(\omega)
+\tilde{\gamma}^{H}(\omega)]}
\notag\\
&\times 
\frac{U_{l\alpha}(\boldQ-\boldq_{1})U_{l^{\prime}\alpha}(\boldQ-\boldq_{1})}
{\omega-\epsilon_{\boldQ-\boldq_{1}}
-i[\tilde{\gamma}(\omega)+\tilde{\gamma}^{H}(\omega)]}.
\end{align}
Since $\Pi_{ll^{\prime}}(\boldQ,\omega)$ for small $Q$ is important 
in analyzing the weak localization~\cite{Bergmann,Nagaoka,Lett}, 
we use the approximations, 
which are appropriate for small $Q$,
\begin{align}
&U_{l\alpha}(\boldQ-\boldq_{1})\sim U_{l\alpha}(\boldq_{1}),
\end{align}
and
\begin{align}
&\epsilon_{\boldQ-\boldq_{1}}
\sim \epsilon_{\boldq_{1}}-\frac{\partial \epsilon_{\boldq_{1}}}{\partial \boldq_{1}}\cdot \boldQ
=\epsilon_{\boldq_{1}}-\boldv_{\boldq_{1}}\cdot \boldQ.
\end{align}
Thus, $\Pi_{ll^{\prime}}(\boldQ,\omega)$ for $\omega > 0$ and small $Q$ is given by
\begin{align}
&\Pi_{ll^{\prime}}(\boldQ,\omega)
\sim 
\sum\limits_{\boldq_{1}}
\frac{U_{l\alpha}(\boldq_{1})U_{l^{\prime}\alpha}(\boldq_{1})}
{\omega-\epsilon_{\boldq_{1}}
+i\tilde{\gamma}(\omega)+i\tilde{\gamma}^{H}(\omega)}\notag\\
&\times 
\frac{U_{l\alpha}(\boldq_{1})U_{l^{\prime}\alpha}(\boldq_{1})}
{\omega-\epsilon_{\boldq_{1}}+\boldv_{\boldq_{1}}\cdot \boldQ
-i\tilde{\gamma}(\omega)-i\tilde{\gamma}^{H}(\omega)}.
\end{align}
In addition, 
we approximate the momentum-dependent $U_{l\alpha}(\boldq_{1})^{2}$ 
and $\boldv_{\boldq_{1}}$ 
as particular values, 
$u_{l\alpha}^{2}=U_{l\alpha}(\boldq_{0})^{2}$ and $\boldv_{\boldq_{0}}$; 
$\boldq_{0}$ is a certain momentum whose magnitude is small. 
This approximation will be appropriate for a rough estimate 
because the main contributions in the sum of $\boldq_{1}$ 
come from the small-$q_{1}$ contributions. 
[We will use the similar approximation to derive 
Eq. (\ref{eq:kappa^VC-simpler}) from Eq. (\ref{eq:kappa^VC-simpler-pre}).] 
As a result of this approximation, 
we can easily perform the sum of $\boldq_{1}$, and 
express $\Pi_{ll^{\prime}}(\boldQ,\omega)$ for $\omega > 0$ and small $Q$ as follows:
\begin{align}
\Pi_{ll^{\prime}}(\boldQ,\omega)
\sim & 
\frac{u_{l\alpha}^{2}u_{l^{\prime}\alpha}^{2}\gamma(\omega)
[1-D_{\textrm{S}}^{H}(\omega)Q^{2}\tilde{\tau}^{\textrm{tot}}(\omega)]}
{\gamma_{\textrm{imp}}[\tilde{\gamma}(\omega)+\tilde{\gamma}^{H}(\omega)]}
,\label{eq:Pi-approx-w>0}
\end{align}
where $\tilde{\tau}^{\textrm{tot}}(\omega)=[\tilde{\gamma}(\omega)
+\tilde{\gamma}^{H}(\omega)]^{-1}
=(\cosh^{4}\theta_{\boldq_{0}}+\sinh^{4}\theta_{\boldq_{0}})^{-1}
[\gamma(\omega)+\gamma^{H}(\omega)]^{-1}$, 
and $D_{\textrm{S}}^{H}(\omega)$ is the spin diffusion constant for $d$ dimensions, 
$D_{\textrm{S}}^{H}(\omega)=\frac{1}{4d}\boldv_{\boldq_{0}}^{2}\tilde{\tau}^{\textrm{tot}}(\omega)$. 
Similarly, 
we obtain the expression of 
$\Pi_{ll^{\prime}}(\boldQ,\omega)$ for $\omega < 0$ and small $Q$,
\begin{align}
\Pi_{ll^{\prime}}(\boldQ,\omega)
\sim & 
\frac{u_{l\beta}^{2}u_{l^{\prime}\beta}^{2}\gamma(-\omega)
[1-D_{\textrm{S}}^{H}(-\omega)Q^{2}\tilde{\tau}^{\textrm{tot}}(-\omega)]}
{\gamma_{\textrm{imp}}[\tilde{\gamma}(-\omega)+\tilde{\gamma}^{H}(-\omega)]}
.\label{eq:Pi-approx-w<0}
\end{align}
Then, 
by using Eqs. (\ref{eq:Pi-approx-w>0}), (\ref{eq:Pi-approx-w<0}), 
and (\ref{eq:BSeq}), 
we can express $\Gamma_{ll^{\prime}}(\boldQ,\omega)$ for small $Q$ as follows:
\begin{align}
&\Gamma_{ll^{\prime}}(\boldQ,\omega)
=  
\frac{\gamma_{\textrm{imp}}^{2}\Pi_{ll^{\prime}}(\boldQ,\omega)}
{1-\gamma_{\textrm{imp}}\Pi_{AA}(\boldQ,\omega)
-\gamma_{\textrm{imp}}\Pi_{BB}(\boldQ,\omega)}\notag\\
\sim &
\begin{cases}
\dfrac{\gamma_{\textrm{imp}}u_{l\alpha}^{2}u_{l^{\prime}\alpha}^{2}\gamma(\omega)}
{\tilde{\gamma}^{H}(\omega)
+\tilde{\gamma}(\omega)D_{\textrm{S}}^{H}(\omega)Q^{2}\tilde{\tau}^{\textrm{tot}}(\omega)}
\ \ \ \ \ \ \ \ \ \ (\omega > 0)\\[8pt]
\dfrac{\gamma_{\textrm{imp}}u_{l\beta}^{2}u_{l^{\prime}\beta}^{2}\gamma(-\omega)}
{\tilde{\gamma}^{H}(-\omega)
+\tilde{\gamma}(-\omega)D_{\textrm{S}}^{H}(-\omega)Q^{2}\tilde{\tau}^{\textrm{tot}}(-\omega)}
\ (\omega < 0)
\end{cases}\hspace{-10pt}.\label{eq:Gamma-last}
\end{align}
This shows that 
$\Gamma_{ll^{\prime}}(\boldQ,\omega)$ does not diverge 
even in the limit $Q\rightarrow 0$ 
because of the damping that is finite only for $H\neq 0$. 
This suggests that
the weak magnetic field suppresses the critical back scattering 
for $\boldQ=\boldq+\boldq^{\prime}$.

We finally analyze the main effect of the weak magnetic field 
on $\kappa_{xx}^{(\textrm{Born})}$ and $\Delta \kappa_{xx}$. 
Substituting Eqs. (\ref{eq:D^R}) and (\ref{eq:D^A}) into Eq. (\ref{eq:kappa^Born}) 
and performing the integral and sums, 
we obtain 
\begin{align}
\kappa_{xx}^{(\textrm{Born})}
\sim 
\frac{1}{TN}
\sum\limits_{\boldq}
\Bigl(
\frac{\partial \epsilon_{\boldq}}{\partial q_{x}}
\epsilon_{\boldq}
\Bigr)^{2}
\Bigl[
-\frac{\partial n(\epsilon_{\boldq})}{\partial \epsilon_{\boldq}}
\Bigr]\tilde{\tau}^{\textrm{tot}}(\epsilon_{\boldq}).\label{eq:kappa^Born-simpler}
\end{align}
In the above calculation, 
we have approximated 
$[-\frac{\partial n(\epsilon)}{\partial \epsilon}]$ 
and $\tilde{\gamma}(\epsilon)+\tilde{\gamma}^{H}(\epsilon)$ 
as $[-\frac{\partial n(\epsilon_{\boldq})}{\partial \epsilon_{\boldq}}]$ 
and $\tilde{\gamma}(\epsilon_{\boldq})+\tilde{\gamma}^{H}(\epsilon_{\boldq})$ 
because the product of the Green's functions in Eq. (\ref{eq:kappa^Born}) 
for $\epsilon > 0$ or for $\epsilon < 0$ 
is large around $\epsilon=\epsilon_{\boldq}$ 
or around $\epsilon=-\epsilon_{\boldq}$, respectively. 
Equation (\ref{eq:kappa^Born-simpler}) shows that 
the change of the lifetime, the inverse of the damping, 
is the main effect of the weak magnetic field on $\kappa_{xx}^{(\textrm{Born})}$. 
Since the lifetime becomes short with increasing $H$, 
the weak magnetic field reduces $\kappa_{xx}^{(\textrm{Born})}$, 
resulting in the positive magneto-thermal-resistance;  
the thermal resistivity is defined as the inverse of the thermal conductivity. 
However, 
this contribution will be small 
because $\tilde{\tau}^{\textrm{tot}}(\epsilon_{\boldq})=
\frac{1}{\tilde{\gamma}(\epsilon_{\boldq})+\tilde{\gamma}^{H}(\epsilon_{\boldq})}
\sim \frac{1}{\tilde{\gamma}(\epsilon_{\boldq})}
[1-\frac{\tilde{\gamma}^{H}(\epsilon_{\boldq})}{\tilde{\gamma}(\epsilon_{\boldq})}+\cdots]$ 
and $\tilde{\gamma}^{H}(\epsilon_{\boldq})/\tilde{\gamma}(\epsilon_{\boldq})$ 
is a small quantity for the weak magnetic field. 
Then, we turn to $\Delta\kappa_{xx}$. 
Since the dominant terms of $\Gamma_{l^{\prime}l}(\boldq+\boldq^{\prime},\epsilon)$ 
in Eq. (\ref{eq:kappa^VC}) 
come from the contributions for small $Q=|\boldq+\boldq^{\prime}|$ 
[see Eq. (\ref{eq:Gamma-last})], 
we set $\boldq^{\prime}=-\boldq$ in Eq. (\ref{eq:kappa^VC}) 
except for $\Gamma_{l^{\prime}l}(\boldq+\boldq^{\prime},\epsilon)$. 
Furthermore, 
for comparison with the result~\cite{Lett} without the magnetic field, 
we introduce the cut-offs for the sum of $\boldq^{\prime}$ in Eq. (\ref{eq:kappa^VC}) 
in the same way as the case without magnetic fields~\cite{Lett}: 
the lower value of $Q=|\boldq+\boldq^{\prime}|$ in the sum 
is replaced by $L^{-1}$, which approaches zero in the thermodynamic limit; 
the upper value of $Q$ is replaced by $L_{\textrm{m}}^{-1}$, 
the inverse of the mean-free path. 
Because of these simplifications, 
Eq. (\ref{eq:kappa^VC}) is reduced to
\begin{align}
\Delta\kappa_{xx}
=&-\frac{1}{TN}
\sum\limits_{\boldq}
\sum\limits_{l_{1},l_{2},l_{3},l_{4}}
e_{l_{1}l_{2}}^{x}(\boldq)e_{l_{3}l_{4}}^{x}(\boldq)
P\hspace{-4pt}\int^{\infty}_{-\infty}\hspace{-2pt}\frac{d\epsilon}{2\pi}\notag\\
&\times 
\Bigl[-\frac{\partial n(\epsilon)}{\partial \epsilon}\Bigr]
\sum\limits_{l,l^{\prime}}
\bar{D}_{l_{4}l^{\prime}}^{(\textrm{A})}(\boldq,\epsilon)
\bar{D}_{ll_{3}}^{(\textrm{R})}(\boldq,\epsilon)\notag\\
&\times 
\bar{D}_{ll_{1}}^{(\textrm{A})}(\boldq,\epsilon)
\bar{D}_{l_{2}l^{\prime}}^{(\textrm{R})}(\boldq,\epsilon)
\sum\limits_{\boldQ}^{\prime}
\Gamma_{l^{\prime}l}(\boldQ,\epsilon),\label{eq:kappa^VC-simpler-pre}
\end{align}
where the prime in the sum of $\boldQ$ represents the cut-offs of the upper and lower values. 
Our theory up to this point is applicable to any dimension; 
hereafter, 
we apply the theory to a two-dimensional case. 
In a similar way to the case for $\kappa_{xx}^{(\textrm{Born})}$, 
we can perform 
the integral and sums in Eq. (\ref{eq:kappa^VC-simpler-pre}). 
As a result, 
we obtain 
\begin{align}
\Delta\kappa_{xx}
&\sim 
-\kappa_{xx}^{(\textrm{Born})}
\dfrac{n_{\textrm{imp}}V_{\textrm{imp}}^{2}}
{8\pi D_{\textrm{S}}^{H}(\epsilon_{\boldq_{0}})}
\tau^{\textrm{tot}}(\epsilon_{\boldq_{0}})
\ln \Bigl(\dfrac{L_{H}}{L_{\textrm{m}}}\Bigr)\notag\\
&= 
-\kappa_{xx}^{(\textrm{Born})}
\dfrac{n_{\textrm{imp}}V_{\textrm{imp}}^{2}}
{[\pi \boldv_{\boldq_{0}}^{2}/(c_{0}^{4}+s_{0}^{4})]}
\ln \Bigl(\dfrac{L_{H}}{L_{\textrm{m}}}\Bigr),\label{eq:kappa^VC-simpler}
\end{align}
where 
$\tau^{\textrm{tot}}(\omega)=[\gamma(\omega)+\gamma^{H}(\omega)]^{-1}$, 
$c_{0}^{4}=\cosh^{4}\theta_{\boldq_{0}}$, 
$s_{0}^{4}=\sinh^{4}\theta_{\boldq_{0}}$, and  
\begin{align}
\hspace{-10pt}
L_{H}=\sqrt{
\dfrac{\tilde{\gamma}(\epsilon_{\boldq_{0}})}{\tilde{\gamma}^{H}(\epsilon_{\boldq_{0}})}
D_{\textrm{S}}^{H}(\epsilon_{\boldq_{0}})
\tilde{\tau}^{\textrm{tot}}(\epsilon_{\boldq_{0}})
}
=L_{\textrm{m}}
\sqrt{\dfrac{\tilde{\gamma}(\epsilon_{\boldq_{0}})}{\tilde{\gamma}^{H}(\epsilon_{\boldq_{0}})}}
.\label{eq:L_H}
\end{align}
In the derivation of Eq. (\ref{eq:kappa^VC-simpler}), 
we have approximated the momentum-dependent $\cosh^{2}\theta_{\boldq}$, $\sinh^{2}\theta_{\boldq}$, 
$\gamma(\epsilon_{\boldq})$, and $\gamma^{H}(\epsilon_{\boldq})$ 
as particular values, 
$\cosh^{2}\theta_{\boldq_{0}}$, $\sinh^{2}\theta_{\boldq_{0}}$, 
$\gamma(\epsilon_{\boldq_{0}})$, and $\gamma^{H}(\epsilon_{\boldq_{0}})$ 
in a similar way to Eq. (\ref{eq:Pi-approx-w>0})
because 
the main contributions in the sum of $\boldq$ in Eq. (\ref{eq:kappa^VC-simpler-pre}) 
come from the small-$q$ contributions 
due to the factor $\frac{\partial n(\epsilon_{\boldq})}{\partial \epsilon_{\boldq}}$. 
$L_{H}$ is a characteristic length of the magnetic-field effect, 
and   
$L_{H}$ is much larger than $L_{\textrm{m}}$ for the weak magnetic field. 
In addition, 
$L_{H}/L_{\textrm{m}}\propto H^{-\frac{1}{2}}$ within the leading order
because 
the leading term of $\tilde{\gamma}^{H}(\epsilon_{\boldq_{0}})$ is proportional to $H$ 
and $\tilde{\gamma}(\epsilon_{\boldq_{0}})$ is independent of $H$. 
From Eq. (\ref{eq:kappa^VC-simpler}),
we can deduce three important properties of the weak-localization correction term:  
one is that the coefficient of the logarithmic dependence of $\Delta \kappa_{xx}$ 
is independent of impurity quantities 
because $n_{\textrm{imp}}V_{\textrm{imp}}^{2}$ in 
$\kappa_{xx}^{(\textrm{Born})}\propto 1/n_{\textrm{imp}}V_{\textrm{imp}}^{2}$ cancels out  
$n_{\textrm{imp}}V_{\textrm{imp}}^{2}$ appearing in Eq. (\ref{eq:kappa^VC-simpler});  
another is that 
$\Delta \kappa_{xx}$ gives a negative contribution to the magneto-thermal-resistance 
because $\ln (L_{H}/L_{\textrm{m}})\propto -\ln H$ within the leading term; 
and the other is that this contribution is not small  
because the coefficient of $\Delta \kappa_{xx}$ is impurity-independent 
and because $\tilde{\gamma}(\epsilon_{\boldq_{0}})/\tilde{\gamma}^{H}(\epsilon_{\boldq_{0}})$, 
appearing in $\ln (L_{H}/L_{\textrm{m}})$, 
is a large quantity for the weak magnetic field. 
Combining Eqs. (\ref{eq:kappa^Born-simpler}) and (\ref{eq:kappa^VC-simpler}), 
we have
\begin{align}
\kappa_{xx}=\kappa_{xx}^{(\textrm{Born})}
\Bigl[
1-\dfrac{n_{\textrm{imp}}V_{\textrm{imp}}^{2}}{[\pi\boldv_{\boldq_{0}}^{2}/(c_{0}^{4}+s_{0}^{4})]}
\ln \Bigl(\dfrac{L_{H}}{L_{\textrm{m}}}\Bigr)
\Bigr].\label{eq:kappa_last}
\end{align}
For the expression without the magnetic field, 
$L_{H}$ in Eq. (\ref{eq:kappa_last}) is replaced by $L$, 
and $\Delta\kappa_{xx}$ gives the negative logarithmic divergence 
in the thermodynamic limit~\cite{Lett}.   
From the arguments in this paragraph, we conclude that 
the negative magneto-thermal-resistance occurs 
in the two-dimensional disordered antiferromagnet 
due to the effect of the weak magnetic field 
on the weak localization. 

\section{Discussion}

We first compare our result with magneto-transport of disordered electron systems. 
As a magneto-transport property of disordered electron systems, 
the longitudinal charge conductivity of electrons, $\sigma_{xx}^{\textrm{C}}$, 
has been often analyzed. 
$\sigma_{xx}^{\textrm{C}}$ in two dimensions 
shows the negative magnetoresistance 
due to the effect of a weak magnetic field 
on the weak-localization correction term 
of $\sigma_{xx}^{\textrm{C}}$~\cite{Nagaoka,Bergmann,Hikami,Maekawa}. 
In a similar way to $\sigma_{xx}^{\textrm{C}}$, 
the longitudinal thermal conductivity of electrons in two dimensions 
may show negative magneto-thermal-resistance. 
This negative magneto-thermal-resistance is similar to our phenomenon. 
However, 
there is at least a major difference between them. 
Since in electron systems a thermal current can induce a charge current, 
magneto-thermal-transport for electrons accompanies magneto-charge-transport for electrons. 
On the other hand, 
our magneto-thermal-transport for magnons 
never accompanies magneto-charge-transport 
because the charge current is absent in magnets, magnetically ordered insulators. 
Because of this major difference, 
our phenomenon will be useful for magneto-thermal-transport free from charge transport. 
In addition to this major difference, 
there is a minor difference: 
the thermal current and energy current are the same in magnets, 
while these are different in electron systems~\cite{Mahan-text}. 

We next discuss implications for experiments. 
First, 
our negative magneto-thermal-resistance will be experimentally observed 
in a quasi-two-dimensional disordered antiferromagnet 
with a weak external magnetic field. 
The more details are as follows. 
Our two-dimensional disordered antiferromagnet (Fig. \ref{fig1}) 
can be experimentally realized 
by replacing part of magnetic ions in a quasi-two-dimensional antiferromagnet 
by different magnetic ions; 
the original magnetic ions and the different ones belong to the same family 
of the periodic table. 
The reasons why we consider such a replacement are 
that magnetic ions in the same family have the same electron number in the open shell, 
resulting in the same $S$, 
and that 
the main difference between different magnetic ions in the same family is 
the difference in the overlap of the wave functions, 
resulting in the difference in the exchange interaction. 
Such an example is a quasi-two-dimensional antiferromagnet in a Cu oxide 
with partial substitution of Ag ions for Cu ions, 
such as La$_{2}$Cu$_{1-x}$Ag$_{x}$O$_{4}$, 
in which Cu ions have a $(3d)^{9}$ configuration 
and Ag ions have a $(4d)^{9}$ configuration~\cite{Lett}. 
In such a quasi-two-dimensional disordered antiferromagnet, 
the weak localization of magnons will be experimentally detectable 
by measuring $\kappa_{xx}$ at a low temperature 
in the absence of an external magnetic field, 
as proposed in Ref. \onlinecite{Lett}. 
If the magnetic field, whose direction is parallel to the directions of the ordered spins, 
is applied to the quasi-two-dimensional disordered antiferromagnet, 
the magnetic-field dependence of $\Delta\kappa_{xx}$ will be 
$\Delta\kappa_{xx}\propto \ln H$ at a low temperature for weak $H$. 
This logarithmic increase is the negative magneto-thermal-resistance 
for the weak localization of magnons in two dimensions. 
Then, 
our negative magneto-thermal-resistance may be useful for enhancing 
the magnitude of the magnon thermal current. 
In addition, 
by utilizing the effects of the weak localization of magnons~\cite{Lett} 
and the weak magnetic field, 
it may be possible to control the magnitude of the magnon thermal current 
in spintronics devices 
because the weak localization is useful for reducing the magnitude, 
and the magnitude can vary by changing the value of the weak magnetic field 
in the presence of the weak localization. 
Since the present possible applications~\cite{Jungwirth,Kajiwara,Uchida,Bauer} 
have focused mainly on non-disordered magnets, 
our previous~\cite{Lett} and present results will provide 
a different possible way for applications using the properties of disordered magnets. 

We finally discuss several directions for further theoretical studies. 
First of all, 
our theory can study the magneto-thermal-resistance in any disordered antiferromagnets 
because this is applicable to disordered antiferromagnets 
for any dimension, any $S$, 
and any lattice with an antiferromagnetic two-sublatice structure.  
As described in Sec. III, 
the equations formulated until Eq. (\ref{eq:kappa^VC-simpler-pre}) 
are applicable to any dimension. 
In addition, 
our theory is applicable even for not large $S$ 
as long as magnons can be defined 
because a ratio of $V_{\textrm{imp}}$ to the magnon dispersion energy 
is independent of $S$ (see Appendix A); 
thus, our theory will be valid 
if temperature is low enough to regard low-energy excitations as magnons.  
Then, 
our theory is useful for studying other magneto-thermal-transport phenomena, 
such as the thermal Hall effect~\cite{ThHall-theory,ThHall} with an external magnetic field, 
in the disordered antiferromagnet. 
While the essential excitations for $\kappa_{xx}$ are intraband, 
the interband excitations are essential for the thermal Hall conductivity~\cite{ThHall-theory}.  
Thus, 
by combining the present result with the result of such a study, 
it is possible to understand the roles of the different kinds of excitations 
in magneto-transport phenomena for disordered magnets.  
Moreover, our theory can be extended to other disordered magnets. 
Such theories may be useful for understanding the roles of the magnetic structure 
in magneto-thermal-transport phenomena 
in the presence of the weak localization of magnons.

\section{Summary}
We have studied the main effect of a weak magnetic field 
on $\kappa_{xx}$ for magnons 
in a disordered two-dimensional antiferromagnet 
in the weak-localization regime. 
We have shown that 
the weak-localization correction term of $\kappa_{xx}$, $\Delta \kappa_{xx}$, increases 
with increasing the magnetic field. 
This increase of $\Delta \kappa_{xx}$ 
is proportional to $\ln H$ within the leading order. 
This phenomenon is negative magneto-thermal-resistance for magnons 
and will be experimentally observed 
in a disordered quasi-two-dimensional antiferromagnet 
in the presence of a weak external magnetic field. 
Our magneto-thermal-transport phenomenon is free from charge transports 
in contrast to the phenomenon for electrons. 
Furthermore, 
our phenomenon may be useful for changing the magnitude of the magnon thermal current 
in antiferromagnetic spintronics devices. 
Then, 
our theory provides a starting point for further studies 
about magneto-thermal-transport phenomena for magnons of various disordered magnets.

\begin{acknowledgments}
This work was supported by CREST, JST, and Grant-in-Aid 
for Scientific Research (A) (17H01052) from MEXT, Japan.
\end{acknowledgments}

\appendix
\section{Derivation of Eq. (\ref{eq:Himp-S})}

In this Appendix, 
we explain how to derive Eq. (\ref{eq:Himp-S}). 
Since its detail has been described in Ref. \onlinecite{Lett}, 
we here describe the main points. 
First, 
we assume that 
substituting part of magnetic ions by different magnetic ions is one kind of disorder, 
and its main effect for disordered Heisenberg antiferromagnets is 
the change of the exchange interaction (see Fig. \ref{fig1}). 
(For the sake of simplicity, we neglect the change of the magnetic anisotropy, 
which is smaller than that of the exchange interaction.) 
With this assumption, 
$\hat{H}_{\textrm{imp}}$ is given by 
\begin{equation}
\hat{H}_{\textrm{imp}}=
2\sum\limits_{\langle \boldi,\boldj \rangle}
\Delta J_{\boldi \boldj}\hat{\boldS}_{\boldi}\cdot \hat{\boldS}_{\boldj},\label{eq:Himp-original} 
\end{equation}
where 
\begin{equation}
\Delta J_{\boldi \boldj}=
\begin{cases}
J^{\prime}\ \ \ (\boldi\in A_{0}, \boldj\in B_{\textrm{imp}})\\
J^{\prime}\ \ \ (\boldi\in A_{\textrm{imp}}, \boldj\in B_{0})\\
J^{\prime\prime}\ \ (\boldi\in A_{\textrm{imp}}, \boldj\in B_{\textrm{imp}})
\end{cases}.
\end{equation}
$A_{0}$ or $B_{0}$ represents $A$ or $B$ sublattice for magnetic ions 
that exist even in the nondisordered system, orange circles in Fig. \ref{fig1}; 
$A_{\textrm{imp}}$ or $B_{\textrm{imp}}$ represents $A$ or $B$ sublattice 
for different magnetic ions, blue circles in Fig. \ref{fig1}. 
Then, we assume that $J^{\prime}$ and $J^{\prime\prime}$ are much smaller than $J$. 
As a result of this assumption, 
the effects of these terms 
on the Neel temperature are negligible, i.e., 
the Neel temperature of our disordered antiferromagnet 
is the same as that of the nondisordered one. 
Since the main terms of $\hat{H}_{\textrm{imp}}$ come from the mean-field type terms, 
we can approximate Eq. (\ref{eq:Himp-original}) as follows:
\begin{align}
\hat{H}_{\textrm{imp}}
=&
-\sum\limits_{\boldi\in A}
V\hat{S}_{\boldi}^{z}
+\sum\limits_{\boldj\in B}
V\hat{S}_{\boldj}^{z}\notag\\
&-\sum\limits_{\boldi\in A_{\textrm{imp}}}
V_{\textrm{imp}}\hat{S}_{\boldi}^{z}
+\sum\limits_{\boldj\in B_{\textrm{imp}}}
V_{\textrm{imp}}\hat{S}_{\boldj}^{z},\label{eq:Himp-next}
\end{align}
where $V=2Sz^{\prime}J^{\prime}$ and $V_{\textrm{imp}}=2Sz^{\prime\prime}J^{\prime\prime}$ 
with $z^{\prime}$ and $z^{\prime\prime}$, 
the coordination numbers for $\Delta J_{\boldi \boldj}=J^{\prime}$ 
and for $\Delta J_{\boldi \boldj}=J^{\prime\prime}$, respectively. 
Due to this expression of $V_{\textrm{imp}}$, 
a ratio of $V_{\textrm{imp}}$ to the magnon dispersion energy 
is independent of $S$. 
In the mean-field approximation for $\hat{H}_{\textrm{imp}}$, 
we have assumed that 
the spin quantum number for impurities is the same as that for magnetic ions of 
the nondisordered system 
because our impurities arise from substituting part of magnetic ions by different magnetic ions 
which belong to the same family in the periodic table 
and because such a substitution does not change the spin quantum number (see Sec. IV). 
In our analyses, 
we neglect the first and second terms of the right-hand side of Eq. (\ref{eq:Himp-next}) 
because their effects in the linear-spin-wave approximation 
are the same as the effect of the magnetic anisotropy of $\hat{H}_{0}$. 
As a result, 
$\hat{H}_{\textrm{imp}}$ is given by Eq. (\ref{eq:Himp-S}). 

\section{Derivation of Eqs. (\ref{eq:kappa-sum}){--}(\ref{eq:kappa^VC})} 

In this Appendix, 
we derive Eqs. (\ref{eq:kappa-sum}){--}(\ref{eq:kappa^VC}) using the linear-response theory 
and a field theoretical technique. 
This derivation is essentially the same as the derivation~\cite{Lett} without 
external magnetic fields; thus, 
we provide the brief explanation below. 
In the linear-response theory, 
$\kappa_{xx}$ is given by 
\begin{align}
\kappa_{xx}=\frac{1}{T}
\lim\limits_{\omega\rightarrow 0}
\dfrac{K_{xx}^{(\textrm{R})}(\omega)
-K_{xx}^{(\textrm{R})}(0)}{i\omega},\label{eq:kappa-start}
\end{align}
where 
\begin{align}
&K_{xx}^{(\textrm{R})}(\omega)=K_{xx}(i\Omega_{n}\rightarrow \omega+i0+),\label{eq:Kxx^R}\\
&K_{xx}(i\Omega_{n})
=\frac{1}{N}
\int^{T^{-1}}_{0}d\tau e^{i\Omega_{n}\tau}
\langle \textrm{T}_{\tau}  
\hat{J}_{\textrm{E}}^{x}(\tau)
\hat{J}_{\textrm{E}}^{x}\rangle.\label{eq:Kxx}
\end{align}
$\Omega_{n}$ is bosonic Matsubara frequency, $\Omega_{n}=2\pi Tn$ 
($n=0,\pm 1, \cdots$). 
Substituting the equation of the energy current operator into Eq. (\ref{eq:Kxx}), 
we can express $K_{xx}(i\Omega_{n})$ in terms of 
the magnon Green's functions in the Matsubara-frequency representation as follows:
\begin{align}
&K_{xx}(i\Omega_{n})
=\frac{1}{N}
\sum\limits_{\boldq,\boldq^{\prime}}
\sum\limits_{l_{1},l_{2},l_{3},l_{4}}
e_{l_{1}l_{2}}^{x}(\boldq)e_{l_{3}l_{4}}^{x}(\boldq^{\prime})
T\sum\limits_{m}
\notag\\
&\times \langle 
D_{l_{4}l_{1}}(\boldq^{\prime},\boldq,i\Omega_{m})
D_{l_{2}l_{3}}(\boldq,\boldq^{\prime},i\Omega_{m}+i\Omega_{n})
\rangle,\label{eq:Kxx-Matsu}
\end{align}
where $D_{ll^{\prime}}(\boldq,\boldq^{\prime},i\Omega_{n})$ 
are the magnon Green's functions before taking the impurity averaging. 
Then, by carrying out the sum of Matsubara frequency in Eq. (\ref{eq:Kxx-Matsu}) 
with a field theoretical technique~\cite{AGD,Eliashberg,NA-Ch} 
and combining that result with Eqs. (\ref{eq:kappa-start}) and (\ref{eq:Kxx^R}), 
we obtain
\begin{align}
\kappa_{xx}=&\frac{1}{TN}
\sum\limits_{\boldq,\boldq^{\prime}}
\sum\limits_{l_{1},l_{2},l_{3},l_{4}}
e_{l_{1}l_{2}}^{x}(\boldq)e_{l_{3}l_{4}}^{x}(\boldq^{\prime})
P\hspace{-4pt}\int^{\infty}_{-\infty}\hspace{-2pt}\frac{d\epsilon}{2\pi}
\Bigl[-\frac{\partial n(\epsilon)}{\partial \epsilon}\Bigr]\notag\\
&\times 
\langle D_{l_{4}l_{1}}^{(\textrm{A})}(\boldq^{\prime},\boldq,\epsilon)
D_{l_{2}l_{3}}^{(\textrm{R})}(\boldq,\boldq^{\prime},\epsilon) \rangle,\label{eq:kappa-RA}
\end{align}
where $D_{l_{4}l_{1}}^{(\textrm{A})}(\boldq^{\prime},\boldq,\epsilon)$ and 
$D_{l_{2}l_{3}}^{(\textrm{R})}(\boldq,\boldq^{\prime},\epsilon)$ 
are   
the advanced and retarded magnon Green's functions 
in the real-frequency representation before taking the impurity averaging. 
In Eq. (\ref{eq:kappa-RA}), 
we have neglected the terms including 
$\langle D_{l_{4}l_{1}}^{(\textrm{R})}(\boldq^{\prime},\boldq,\epsilon)
D_{l_{2}l_{3}}^{(\textrm{R})}(\boldq,\boldq^{\prime},\epsilon)\rangle$ 
and $\langle D_{l_{4}l_{1}}^{(\textrm{A})}(\boldq^{\prime},\boldq,\epsilon)
D_{l_{2}l_{3}}^{(\textrm{A})}(\boldq,\boldq^{\prime},\epsilon)\rangle$ 
because those are higher-order contributions 
in the weak-localization regime~\cite{Nagaoka,Lett}. 
Then, by using the perturbation expansion of $\hat{H}_{\textrm{imp}}$ 
in Eq. (\ref{eq:kappa-RA}), 
we can take the impurity averaging. 
As a result, 
we can express $\kappa_{xx}$ in the weak-localization regime 
as Eqs. (\ref{eq:kappa-sum}){--}(\ref{eq:kappa^VC}).


\end{document}